# The Order and Integration of Knowledge

*Moorad Alexanian*

*William Oliver Martin published "The Order and Integration of Knowledge" in 1957 to address the problem of the nature and the order of various kinds of knowledge; in particular, the theoretical problem of how one kind of knowledge is related to another kind. Martin characterizes kinds of knowledge as being either autonomous or synthetic. The latter are reducible to two or more of the autonomous (or irreducible) kinds of knowledge, viz., history (H), metaphysics (Meta), theology (T), formal logic (FL), mathematics (Math), and generalizations of experimental science (G). Metaphysics and theology constitute the two domains of the ontological context while history and experimental science are the two domains of the phenomenological context. The relation of one kind of knowledge to another may be instrumental, constitutive, and/or regulative. For instance, historical propositions are constitutive of G, metaphysical propositions are regulative of G, and propositions in formal logic and mathematics are instrumental to G. Theological propositions are not related to G and so there is no conflict between science and theology. Martin's work sheds light on the possible areas of incompatibility between science and religion.*

# Preamble

The finite nature of the human mind is evident by the need to understand reality by a process of analysis. This process of taking things apart has resulted in a multitude of disciplines as manifested in the existence of many departments in our institutions of higher learning.  It is clear that each autonomous kind of knowledge deals primarily with a certain aspect of reality and as such it is based on a specific type of evidence that are used to establish the truth or falsehood of given propositions in that field.  For instance, it is foolish for a scientist to require the same kind of evidence, which is appropriate to establish truthful statements in the experimental sciences, from a theologian who studies the intrinsic nature of God and how He interacts with His creation. The theologian has his own source of evidentiary data that differs from that of the scientist. It is important to remark that the data used in all kinds of knowledge is made up of unique historical events represented by historical propositions that are based on sensations, perceptions, memories, and extant records of past events.  Therefore, human reasoning, using essentially historical data stored in the brain, produces human knowledge and understanding.  It is interesting that C. S. Lewis indicates, "The rational and moral element in each human mind is a point of force from the Supernatural working its way into Nature."[1] Therefore, human reasoning is not entirely interlocked with the physical aspect of Nature. This is equivalent to the statement of René Descartes that "matter cannot think."[2]

# Autonomous and Synthetic Kinds of Knowledge

What are the basic, autonomous kinds of knowledge needed to analyze and comprehend the whole of reality?   Martin[3] considers as autonomous the following kinds of knowledge, history (H), metaphysics (Meta), theology (T), formal logic (FL), mathematics (Math), and generalizations of experimental science (G). Metaphysics and theology constitute the two domains of the ontological context and the others, viz., H, FL, Math, and G, as positive kinds of knowledge. History and experimental science are the two domains of the phenomenological context whereas formal logic is the domain of intentional context and mathematics that of formal context.  In addition, Martin[3] considers synthetic kinds of knowledge those that result from the integration of a positive kind of knowledge with the ontological (metaphysics and/or theology). For instance, the human-social sciences, notably, anthropology, culturology, economics, philology, psychology, and sociology are all examples of synthetic kinds of knowledge that integrates scientific studies with the humanities, which fundamentally deals with the nature of humans as deduced from history. Similarly, the philosophy of history, of mathematics, of science, or of Nature, which are synthetic kinds of knowledge, the mode or aspect studied is integrated with the mode of existence or being, which is the subject matter of metaphysics and theology. Note that incompatibilities in the synthetic kinds of knowledge are reducible to those among the autonomous kinds of knowledge that constitutes them. In a sense, Martin systemized the epistemology and metaphysics of Jacques Maritain[4] who dealt with the intrinsic diversity and nature of knowledge that is rational and speculative, philosophical and scientific.

"Now, whenever a science is considered from the standpoint of the mode studied, and not from the standpoint of the being that the mode is of, then that science is considered in its *positive* sense.  A positive science is one which is defined in abstraction from the metaphysical because it is concerned with a mode *qua* mode and not with a mode of being.  Experimental science is positive in this sense.  This does not mean that it is anti-metaphysical; it is simply non-metaphysical.  What is anti-metaphysical is positivism, which is the position that knowledge is limited to a 'mode,' and that there can be no knowledge of a 'mode of being' because there is no science of being *qua* being.  Experimental science is positive science, but it is not the same as positivism.  Positivism is a doctrine *about* it, and hence is not *of* it.  A proposition about experimental science is not necessarily a proposition of it."[5] Therefore, one must distinguish between the causes supposed in the model descriptions of different aspects of reality and the actual causes. The former may be described as secondary while the latter as primary, e.g., God being the primary cause as proposed by Nicolas Malebranche.[6] This may suggest an understanding of René Descartes' notions of "universal and primary cause" and "secondary and particular causes."[7] God is the primary cause of the actual, moment-by-moment temporal development of all that is; whereas, the secondary causes are those that we ascribe to the models that we construct of all that there is, which are based on our sensations, perceptions and memories.  Our understanding of Descartes' primary causes is in agreement with the "occasionalism" of Malebranche and is contrary to the view of Tad Schmaltz who considers creatures rather than God as the casual source of natural change rather than merely "occasional causes."[8]

## Nature of Human Knowledge

Knowledge is summarized in propositions. For instance, historical propositions refer to that which is factually known (potentially or actually) of the process of historical events both human and natural past events. Historical propositions are instrumental in a general sense to propositions of metaphysics, formal logic, and mathematics.[9]  Scientific laws of Nature are generalization of historical propositions, experimental science G, where the data expressed in historical propositions can be obtained, in principle, by purely physical devices.[10] Therefore, history H in the form of historical propositions is wholly constitutive of experimental science G.  The subject matter of formal logic (FL) and mathematics (Math) are the nonphysical, mental or mathematical constructs of the "real things," which are based on data detected by human senses and/or purely physical devices. Human rationality develops formal logic and creates mathematics to summarize data into laws of Nature that lead to theoretical models covering a wide range of phenomena. However, scientists deal with secondary causes. First causes involve metaphysical (ontological) questions, which regulate science. Without the ontological, neither the generalizations nor the historical propositions of the experimental sciences would be possible. The integration of all the positive sciences with the ontological gives us the "real thing" that actually exists. Metaphysics and theology, the fields that encompass the knowledge of being, constitute the two domains of the ontological context that deal with the mode or aspect including the existence or being of the "real thing."  There is a nested sequence of mental abstractions and constructions in the human mind organizing and making sense of the reality based on physical data obtained by purely physical devices and data obtained by humans as "living detectors" of the physical, nonphysical, and the supernatural aspects of Nature.[11]

## Relations between Autonomous Kinds of Knowledge

In Table 1, the relations between the different autonomous or irreducible kinds of knowledge are specified[12] in terms of, "constitutive of" (Con), "instrumental to" (Inst), and "regulative of" (Reg).  For instance, metaphysics (Meta) is regulative of all the different kinds of knowledge, except theology (T) where metaphysics (Meta) is constitutive of some theological propositions. The latter means that some theological proposition $t_1$ implies a particular metaphysical proposition $m_1$, viz., "If $t_1$, then  $m_1$." The converse, "If $m_1$, then $t_1$," does not follow since otherwise the theological proposition $t_1$ is constitutive of the metaphysical proposition $m_1$ and so the truth of $t_1$ would be necessary for the truth of $m_1$. The constitutive aspect of historical propositions for the generalizations of experimental sciences (G) means that given the set of historical propositions $\{h_i\}$, $(i = 1, 2, …., n)$ and the generalization of them by $g_1$, one has that $g_1 \to (h_1, h_2, …., h_n)$. Therefore, the set $\{h_i\}$ is wholly constitutive of $g_1$. The induction or inference, if $(h_1, h_2, …., h_n) \to$ (probably) $g_1$, is the *modus operandi* of the experimental sciences where probability theory is used to indicate which of a given set of generalizations $\{g_1, g_2, …\}$ is most likely to be true in the light of the data and any other evidence at hand.[13]

Mathematical propositions are instrumental in discovering and summarizing the generalizations, which are constituted by the historical propositions. Therefore, the generalization $g_1$, sometimes codified in a theoretical model, gives rise to predictions, say, $g_1 \to h_P$. If $h_P$ is false, then $g_1$ is also false and so the generalization or underlying theory is falsified. "Facts (historical propositions) are as relevant to metaphysics as to experimental science, but not in the same way; for they are instrumental to the discovery of metaphysical truth, but are constitutive as evidence of the generalizations of experimental science."[14]

TABLE 1

Relations between autonomous kinds of knowledge*

|  | H | Meta | T | FL | Math | G |
|---|---|---|---|---|---|---|
| **H** | X | Inst | Con of Some | Inst | Inst | Con |
| **Meta** | Reg | X | Con of Some | Reg | Reg | Reg |
| **T** | None | None | X | None | None | None |
| **FL** | Inst | Inst | Inst | X | Inst | Inst |
| **Math** | Inst | Inst | Inst | Inst | X | Inst |
| **G** | Con of Some | Inst | Inst | Inst | Inst | X |

*The order of knowledge of historical propositions (H), metaphysical propositions (Meta), theological propositions (T), formal logic propositions (FL), mathematical propositions (Math), and the generalizations of experimental science (G) together with their interrelationships described by "instrumental to" (Inst), "regulative of " (Reg), and "constitutive of "(Con).

The analysis of elements of reality into its different aspects, viz. physical, nonphysical, or supernatural, gives rise to the different kinds of knowledge needed to give a true description and understanding of that which is detectable by purely physical devices and by humans as "detectors."[11] The integration of all kinds of knowledge is the object of metaphysics, which delimits the possible and is regulative of all the positive kinds of knowledge, viz., H, FL, Math, and G, and is partially constitutive of T (See Table 1). Historical propositions dealing with the physical aspect of Nature are constitutive of the generalizations of experimental science and form the basis for unadulterated science and the discovery of the laws of Nature. These generalizations of physical data into laws, say $(g_1, g_2, …, g_{10})$, in turn imply a minimal metaphysics $m_1$, dictated by some sort of Ockham's razor, which forms a foundation of our understanding of the physical aspect of Nature. Symbolically, one has $(g_1, g_2, … g_{10}) \to m_1$, where metaphysics $m_1$ deals merely with the physical aspect of Nature. Edwin Thompson Jaynes has written extensively on probability theory as the logic of science.[13]

## Scientific and Theological Metaphysics

An example of an implied scientific metaphysics $m_1$ is the principle that Nature can be understood (hypothesis of comprehensibility) and exclusion of or dispensing with the cognizing subject (objectivation).[15] It is clear that any metaphysics $m_2$, which contains metaphysics $m_1$ as a subset, is equally compatible with the generalizations ($g_1, g_2, ..., g_{10}$). The stronger constitutive relation of metaphysics to some theologies means that given theology $t_1$, then metaphysics $m_3$ must be true, viz., $t_1 \rightarrow m_3$. Theology and metaphysics constitute the two domains of the ontological context of the whole of reality, i.e., the physical, nonphysical and supernatural aspects of Nature.[11] Since experimental science (G) is concerned only with the physical aspect of Nature, a possible incompatibility between experimental science and theology, if any, would be in the physical aspect only. Metaphysics $m_1$ cannot contain metaphysics $m_3$ as a subset since the subject matter of metaphysics $m_1$ is the domain of the phenomenological context and thus regulative of only the purely physical. Therefore, a theistic worldview would be based on the metaphysics $m_4$, which is the union of $m_1$ and $m_3$, that is, $m_4 = m_1 \cup m_3$. To insist of the exclusivity of metaphysics $m_1$ would correspond to a form of physicalism or materialism and thus the elimination of theology $t_1$. This is a form of reductionism, which violates the order of knowledge.

The laws of experimental science are quite consistent with most theological propositions. It is in the study of unique historical events—say, in cosmological or biological evolution—where the conflict between science and religion may arise. Religion, as a kind of knowledge, is a synthetic or reducible kind and is constituted by several of the autonomous or irreducible kinds of knowledge listed in Table 1. Experimental science *qua* generalization of historical propositions has nothing whatsoever to say regarding a particular historical proposition that is not in the class of historical propositions that gave rise to the particular generalization. In other words, if $g_1 \rightarrow (h_1, h_2 ....., h_n)$, where the parentheses denote a class of historical propositions, then one cannot conclude that the particular historical proposition $h_{n+1}$ is not possible owing to the generalization $g_1$, which is based on the class of historical propositions ($h_1, h_2 ....., h_n$). In particular, the results of experimental science cannot be used to disprove the possible existence of miracles as unique, historical events. For instance, in 1931 Paul Dirac showed that the existence of magnetic monopoles lead to the quantization of electric charge, a very fundamental feature of Nature whereby the electric charge of any object, other than quarks which are fractionally charged, is an integer multiple of the charge of the electron.[16] Blas Cabrera presumably detected the hypothetical magnetic monopole.[17] So far, however, this remains the only experimental detection and so it may be that magnetic monopoles do exist and the event, whatever it was, has not been repeated. Therefore, in the order of being, the historicity of the magnetic monopole is on a par with that of a miracle owing to its historical uniqueness.

## Regulative Character of Metaphysics

The regulative character of metaphysics for the experimental sciences and the other positive sciences, viz., formal logic (FL), mathematics (Math), and history (H), follows from the truth table of logical implications, Table 2. Let the following implications be true for the experimental generalizations $\{g_1, g_2, ..., g_{20}\}$ and ($g_1, g_2, ..., g_{10}$) and the metaphysical propositions $\{m_1, m_2\}$, viz., ($g_1, g_2, ..., g_{10}$) $\rightarrow m_1$ and ($g_{11}, g_{12}, ..., g_{20}$) $\rightarrow m_2$. If $m_1$ and $m_2$ are incompatible, say $m_1$ is true and $m_2$ is false, then by Table 2 the set

($g_{11}$, $g_{12}$, …, $g_{20}$) is false. Therefore, metaphysics is regulative of the positive sciences. Metaphysics is actually partially constitutive of theology, which is a stronger relation than metaphysics being merely regulative of theology. Note that the meaning of "regulative" is such that an autonomous kind of knowledge A can be constitutive of another autonomous kind of knowledge B if, in fact, A is not regulative of B. Therefore, as one domain of the ontological context, metaphysics is partially constitutive of the other domain, theology.

TABLE 2

**L**ogical Implication*

| p | q | p → q |
|---|---|---|
| T | T | T |
| T | F | F |
| F | T | T |
| F | F | T |

*T = true, F = false

# Physical Aspect of Nature

Erwin Schrödinger, founder of quantum mechanics together with Werner Heisenberg, was the forerunner in evolutionary biology, genetics, and indeed a great philosopher.[18] P. A. M. Dirac remarked that Schrödinger's equation underlies "a large part of physics and the whole of chemistry."[19] Schrödinger was puzzled by the agreement of the existence of a common, real world observed by two different observers. "Each person's sense-world is strictly private and not directly accessible to anyone else, this agreement is strange, what is especially stranger is how it is established."[20] Schrödinger asked, "How do we come to know of this general agreement between two private worlds, when they admittedly are private and always remain so?"[21] Concerning his holistic view of Nature, Schrödinger considers two hypotheses.[22] Schrödinger further indicates, "I have therefore no hesitation in declaring quite bluntly that the acceptance of a really existing material world, as the explanation of the fact that we all find in the end that we are empirically in the same environment, is mystical and metaphysical."[23] Now consciousness is a moment-by-moment awareness of our temporal existence and surroundings. Thus, human knowledge has access only to snapshots and flashbacks of reality. In Christian theology, God is the being forever conscious and thus eternal that does not exist in time. God has no history and so He experiences the whole of reality as an eternal "present for Him."[24] The metaphysical and mystical nature that Schrödinger ascribes to different spheres of consciousness that recognize that we all live in the same world can partially be demystified if one considers the objective nature of scientific data that is collected with the aid of purely physical devices. Of course, the metaphysical and mystical nature of consciousness, rationality and even life itself remains.

C. S. Lewis seeks answers to the same fundamental questions regarding the acquisition of knowledge by humans. In the process of studying the notion of miracles, their possible occurrence, and their supernatural nature, Lewis makes a detailed analysis of what Nature is and what is the nature of human rationality and morality. "If our argument has been sound, rational thought or Reason is not interlocked with the great interlocking system of irrational events which we call Nature."[25] In particular, "Hence every theory of the universe which makes the human mind a result of irrational causes is inadmissible, for it would be a proof that there are no such things as proofs. Which is nonsense."[26] The laws of Nature govern the physical aspect of Nature and thus possess no elements of free will or rationality. It is in this sense, that the behavior of Nature is understood by Lewis to be irrational. Of course, there may be intelligence behind the workings of Nature, as in the Christian faith where God not only created the whole of reality but He also sustains the creation moment-by-moment into a continuous state of existence.

It interesting that Schrödinger posits the need of the metaphysical and the mystical, whereas Lewis considers rationality as necessary elements to understand natural, spatiotemporal events. These are actually the assumptions of comprehensibility and objectivation considered by Schrödinger.[27] Lewis further indicates human reason and morality as proofs of the supernatural.[28]

One can then surmise that conscious, rational humans develop the theories of the workings of the physical aspect of Nature. However, human rationality is not present in the physical but forms part of the nonphysical aspect of humans--self, consciousness, and rationality--that set up experiments and develop theories of the physical aspect of reality. It is inherently the differing theological presuppositions of Schrödinger and of Lewis that the higher order ontological inferences of Schrödinger ends with the Upanishads and that of Lewis with the Bible. In particular, "why our perceiving and thinking self is nowhere to be found *within* the world-picture: because it itself *is* this world-picture."[29] Schrödinger does not ascribe individuality or distinctness to self whereas Lewis does.[30]

The fundamental question in science, and for that matter, rational thinking in general, is how to decide between differing available models or hypotheses in order to account for the existing data. Of course, one has to be rather clear on what data is being considered and how the data is collected in that particular kind of knowledge. One best way to avoid confusion is to characterize a kind of knowledge according to a subject matter as Martin does.[3] The question is how one does that in an unambiguous, operational fashion so that it is clear what kind of evidence would be necessary in order to establish the truth or falsehood of claims being made. Of course, what prior information is used to analyze such data is crucial. In order to keep track of what assumptions are being made in the process of developing scientific theories of Nature, one must specify what the data D is that one is taking into account and what prior information X one is assuming to analyze the data.[31] Of course, the division of what is data and what is prior information is a purely logical one. In any case, any additional information beyond that provided by the data of the current problem is by definition prior information. Bayesian methods are used whereby the mathematical rules of probability provide consistent rules for conducting inferences.[31] Einstein[32] indicates that physical influences can propagate only forward in time (causality). However, logical connections, which may or may not correspond to causal physical influences, propagate equally well in either direction, viz., "one man's prior probability is another man's posterior probability."[33] In Bayes' theorem, the logical product, say, HX denotes the proposition that both propositions H and X are true. Therefore, in order to know the likely truth of H, viz., the posterior probability $P(H|DX)$, given the data D and prior X, one needs not only the sampling distribution $P(D|HX)$ but also the prior probabilities $P(D|X)$ for D and $P(H|X)$ for H. All probabilities are necessarily conditional on the prior information X since by Bayes' theorem[33] $P(H|DX) = P(H|X) P(D|HX)/P(D|X)$. If parts of X are irrelevant to the problem at hand, then such parts of X will cancel out mathematically and so $P(H|DX)$ will not depend on such parts of X and so are truly irrelevant in determining the likely truth of H. It is clear, for instance, that theological considerations included in the prior information assumed by Isaac Newton that led to his theory of gravitation are irrelevant in relating the data logically to the model or hypothesis.

## Science and Reality

The term "science" is equivocal since it can mean, (a) the method of arriving at generalization of historical propositions that constitute G, (b) the mass of information, accrued by such methods, and (c) the theories developed to summarize the data into laws of Nature. In addition, one needs to distinguish between the experimental sciences from the observational sciences, say astronomy, paleontology, etc., and the historical sciences, say forensic science, cosmology, evolution of life on Earth, etc. Here the term "science" is defined by its subject matter, viz., the physical aspect of the whole of reality.[11] Thus, "science aims at nothing but making true and adequate statements about its object."[34] Therefore, the principle of objectivation, which together with the principle of understandability of Nature form the basis of the scientific method[34], is accomplished here by considering the subject matter of science data that can be collected, in principle, by purely physical devices thus achieving objectivity. Therefore, the laws of experimental science are generalizations of historical propositions --that is, experimental data--thus all physical laws are based on statistics.[13] Note that consciousness and rationality are purely nonphysical, since purely physical devices cannot detect them. In addition, life cannot be reduced to the purely physical, so living beings are both physical and nonphysical. However, despite the difficulty of reducing life to the purely physical, it is interesting that Schrödinger considers a genuinely physical--rather than nonphysical, not to say, supernatural--law in order to interpret life by the ordinary laws of physics.[35] Schrödinger invokes the 1914 work of Max Planck on dynamic and statistical regularity, which makes a fundamental distinction between reversible and irreversible processes, viz., "order-from-order" and "order-from-disorder." Accordingly, Schrödinger's new principle for the understanding of life is nothing new to physics but the "order-from-order," which is the same that governs large dynamical systems, say the motion of planets or clocks. Of course, it is not at all clear how the "order-from-order" that already exists in living beings emerges from the dynamical "order-from-order" governing purely physical systems.

With regard to the notion of "science," an important issue is if a clear demarcation can be made between science and nonscience.[36] If such a demarcation exists, then the question what science is would have an unequivocal answer. It is interesting that many authors use the term "science" but do not quite define it, but do refer to the scientific method, "which has a proven track record dating back at least to Francis Bacon and which has been embellished by modern philosophers of science, particularly by Karl Popper's criterion of falsifiability."[37] For instance, Keith Thomson[37] mentions, in particular, Ronald L. Numbers, Kenneth R. Miller, Alvin Plantinga, Lawrence M. Krauss, and Robert Wuthnow all contributor essayists in the Terry Lectures on the religion and science debate who do not define the term "science."[37] Thomson considers the old Oxford English Dictionary definition of science "in terms of a method of enquiry, applied to organized knowledge, leading to the discovery of general laws, and restricted to those branches of study that relate to the phenomena of the material universe and their laws."[37] Our definition of science is based on its subject matter, which is the physical aspect of Nature.

Despite not providing a clear definition of what is science, Alvin Plantinga claims, "There is a superficial conflict but deep concord between science and theistic religion, but superficial concord and deep conflict between science and naturalism."[38] Plantinga takes naturalism to be the thought that there is no such person as God, or anything like God, and so naturalism is a strictly atheistic worldview. Of course, a generous description of the whole of reality would span, in the order of complexity, from the purely physical to the supernatural.[10] These two extreme aspects of reality are bridged by a nonphysical realm, which would include elements of life, man, consciousness, rationality, mental and mathematical abstractions, etc.

The demarcation between science and nonscience can be made definite if one adopts the definition of science not only as the study of the physical aspect of Nature but with the further proviso that the data that makes up the subject matter of science is that which can be obtained, in principle, solely by purely physical devices.[10] In fact, the term "scientist" was introduced in the nineteenth century when natural philosophy became a synonym for physics and science.[39] In the words of Schrödinger, "The strange fact that on the one hand all our knowledge about the world around us, both that gained in everyday life and that revealed by the most carefully planned and painstaking laboratory experiments, rests entirely on immediate sense perception, while on the other hand this knowledge fails to reveal the relations of the sense perceptions to the outside world, so that in the picture or model we form of the outside world, guided by our scientific discoveries, all sensual qualities are absent."[40] It is interesting that Democritus of Abdera clearly understood this state of affairs already in the fifth century B.C. prior to the advent of the sophisticated instrumentations of today.[41]

Keith Ward indicates that the definition of science is subject of much debate but enumerates the main characteristics of natural science, "data are publicly observable, measurable, repeatable, and agreed on by all competent observers."[42] He also raises the important question, "Are there any data that are not publicly observable, measurable, repeatable, and agreed on by all competent observers?"[42] Of course, Ward is alluding to experiences in religion, "Overall, it seems that distinctive forms of experience that may be termed *religious* are important to religion and that, without them, religions might not have come into existence."[43] Clearly the subject matter of science and religion, the physical vs. the nonphysical (theology), differ and thus the former can be detected by purely physical devices whereas the latter requires humans as "detectors."

## Compatibility of Science and Religion

In his seminal 1989-1991 Gifford Lectures on the place of religion in an age of science, in particular, the compatibility of science and religion, Ian Barbour[44] sets the stage for the fruitful dialogue between science and religion. Barbour deals, amongst many topics, with the success of the methods of science and the danger of supposing that science may be the only reliable path to knowledge. In addition, the present scientific description of Nature has brought about a radical new view of Nature mainly by the counterintuitive views provided by quantum physics and relativity, which can be made to impinge on the nature of life and mind. Barbour considers, how all these discoveries affect, in particular, the doctrine of human nature and the doctrine of creation? The question of what is reality comes to the fore when dealing with these fundamental questions. Barbour develops a "relational and multileveled view of reality. In this view, interdependent systems and larger wholes influence the behavior of lower-level parts. Such an interpretation provides an alternative to both the classical dualism of spirit and matter (or mind and body) and the materialism that often replaced it."[45] The whole of reality considered in our work is physical/nonphysical/supernatural.

It is difficult to discuss the question of the compatibility or the conflict between science and religion without a clear understanding of the meaning of these terms. Of course, the word "religion" is quite equivocal and unless clearly defined will easily lead to all sorts of misunderstandings. The broadness of the meaning of the word "religion" can be gauged by the following, "Religion, whether understood in broad cultural terms or in more narrow theological categories, reflects a search for meaning and unity, for wholeness and relatedness."[46] Religion is a synthetic kind of knowledge that includes several autonomous kinds of knowledge, for instance, theology (T), history (H), etc. There is no conflict between science and theology and so if a conflict or incompatibility exists between science and religion, it is between science and some of the autonomous kinds of knowledge that constitutes religion other than theology. For instance, the Christian faith is based essentially on the historicity of Jesus of Nazareth, his death, and his resurrection. Absent those historical events, there would be no Christian faith. Therefore, the Christian faith as a religion has essential historically unique elements in its constitution in addition to the theological propositions that underlie the supernatural aspect of the faith. Therefore, there is no incompatibility between the Christian faith and the experimental sciences that underlie the laws of Nature, which are generalizations of historical propositions and do not apply to unique, historical events.

Robert Wuthnow observes that, "science and religion come into conflict because neither stays neatly in its respective sphere."[47] But, what are the natures of the spheres of influence of science and religion? Alvin Plantinga indicates, "There is an alleged conflict between the *epistemic attitudes* of science and religion. The scientific attitude, so it is said, involves forming belief on the basis of empirical investigation, holding belief tentatively, constantly testing belief, and looking for a better alternative; the religious attitude involves believing on faith."[48] However, is this a criticism of religion or actually a comment on the differing nature of science and religion according to the kinds of knowledge they are as discussed above?

Barbour finds methodological parallels between science and religion indicating that science is not as objective, nor religion as subjective as has been claimed, "Scientific data are theory-laden, not theory-free. Theoretical assumptions enter the selection, reporting, and interpretation of what are taken to be data. Moreover, theories do not arise from logical analysis of data but from acts of creative imagination in which analogies and models often play a role. Conceptual models help us to imagine what is not directly observable."[49] Barbour further observes, "Many of these same characteristics are present in religion. If the data of religion include religious experience, rituals, and scriptural texts, such data are even more heavily laden with conceptual interpretations. In religious language, too, metaphors and models are prominent."[49] Finally, Barbour contrasts the similarities and differences between science and religion, "Clearly, religious beliefs are not amenable to strict empirical testing, but they can be approached with some of the same spirit of inquiry found in science. The scientific criteria of coherence, comprehensiveness, and fruitfulness have their parallels in religious thought."[49] The characterization of different kinds of knowledge and the data needed to make truth statements helps make sense of the whole of reality, viz., physical/nonphysical/supernatural, and allows finding the source of actual and apparent incompatibilities between science and religion.

## Discussion and Conclusion

The kinds of knowledge needed to study and know all that exists are characterized here by their subject matter. This allows us to create order and eventually integrate all the different kinds of knowledge needed to know truly the real existing things. The question, what the whole of reality is and how do we obtain data for it is addressed within the context of the Christian worldview that considers a human being as body, mind, and spirit (soul), which is consistent with Cartesian ontology of only three elements: matter, mind, and God. This indicates what the whole of realty is and how do we obtain data for it. The unequivocal definition of science, as the study of the physical aspect of reality, strictly defines the actual subject matter of physics. However, reductionists use such a definition to suppose that science encompasses the whole of reality and is the only means of knowing. Clearly, we do not adhere to physicalism since purely physical devices, which in principle collect the data of the physical aspect of Nature, do not detect the aspects of humans that are actually nonphysical. Note that reductionists consider nonphysical aspects of humans as actually emerging and explicable from the purely physical. Therefore, the study of humans that goes beyond the physical aspect and ventures into the nonphysical/supernatural is tricky, owing to the difficulty of obtaining unambiguous and consistent data. Note that in biology, psychology, sociology, neuroscience, economics, etc., one is relying more and more on a quantifiable description of humans, this is tantamount to emphasizing the physical over the more important nonphysical aspects of humans. The Bible deals with humans in historical contexts, which are not amenable to generalizations into scientific laws. In fact, the importance of the Bible is the truth it provides of the nonphysical/supernatural aspects of humans. Therefore, knowledge of the physical aspect of Nature tells us nothing of the true nature of humans and, least of all, of the nature of God.

Schrödinger's scientific metaphysics, viz., the hypothesis of comprehensibility and objectivation, is compatible with the metaphysics implied by theology. The metaphysics underlying science does not regulate all means of knowing and so there can be no conflict between science and theology. In fact, the subject matter of science and the content of the Bible overlap only in the physical aspect of Nature, since Nature itself is a physical/nonphysical/supernatural entity owing to the existence of humans. The Bible deals with ontological, rather than experimental issues since science deals with the phenomenological context and theology and metaphysics constitute the two domains of the ontological context of the whole of reality. Therefore, only apparent incompatibilities can arise between the experimental sciences and theology with regard to the physical aspect of reality. Herein lies the only possible source of a conflict between the generalization in the experimental sciences and theology. Of course, with regard to the conflict between science and religion, the synthetic nature of religion as a kind of knowledge augments the possible areas of conflict considerably and it is hoped that the work presented here helps identify and clarify such areas where incompatibilities may arise.

## Notes

[1] C. S. Lewis, *Miracles: A Preliminary Study* (New York: The Macmillan Company, 1971), 40.

[2] Margaret Dauler Wilson, *Descartes* (New York: Taylor & *Francis*, 2005), 159.

[3] William Oliver Martin, *The Order and Integration of Knowledge* (Ann Arbor: The University of Michigan Press, 1957).

[4] Jacques Maritain, *An Introduction to Philosophy* (New York: Sheed and Ward, 1962), 67. See, also, Jacques Maritain, *The Range of Reason* (New York: Charles Scribner's Sons, 1952), 3-18.

[5] Martin, *The Order and Integration of Knowledge*, 175-6.

[6] Steven Nadler, "Malebranche on Causation," in *Malebranche*, ed. Steven Nadler (Cambridge: Cambridge University Press, 2000), 112-38.

[7] Tad M. Schmaltz, *Descartes on Causation* (New York: Oxford University Press, 2008), 90.

[8] Ibid., 4.

[9] Martin, *The Order and Integration of Knowledge*, 319.

[10] Moorad Alexanian, "Debate about science and religion continues," *Physics Today (Letter)* 60, no. 2 (2007), 10 &12.

[11] Moorad Alexanian, "Physical and Nonphysical Aspects of Nature," *Perspectives on Science and Christian Faith* 54, no. 4 (2002): 287-8.

[12] Martin, *The Order and Integration of Knowledge*, 318.

[13] E. T. Jaynes, *Probability theory: The Logic of Science* (New York: Cambridge University Press, 2003).

[14] Martin, *The Order and Integration of Knowledge*, 122.

[15] Erwin Schrödinger, *What is Life? & Other Scientific Essays* (Garden City: Doubleday Anchor Books, 1956), 182-3.

[16] P.A.M. Dirac, "Quantized Singularities in the Electromagnetic Field," *Proc. Roy. Soc. London* A 133 (1931), 60-72.

[36]Gregory Peterson, "The Scientific Status of Theology: Imre Lakatos, Method and Demarcation," *Perspectives on Science and Christian Faith* **50** (March 1998): 22-31.

[37]Keith Thomson, "Introduction," in *The Religion and Science Debate: Why Does It Continue?* ed. Harold W. Attridge (New Haven, CT: Yale University Press, 2009), 4.

[38]Alvin Plantinga, *Where the Conflict Really Lies: Science, Religion, and Naturalism* (New York, NY: Oxford University Press, 2011), ix-xiv.

[39] "By extending the application of the term natural philosophy to the mathematical sciences, most notably mathematical physics, Newton may have begun a tradition that reached fruition in the nineteenth century when natural philosophy came to be called physics, and, often enough, science in general." Edward Grant, *A History of Natural Philosophy* (New York: Cambridge University Press, 2007), 316.

[40]Erwin Schrödinger, *What is life? with Mind and Matter & Autobiographical Sketches* (Cambridge, Great Britain: Cambridge University Press, 1967), 153.

[41]Ibid., 163.

[42]Keith Ward, *The Big Questions in Science and Religion* (West Conshohocken, PA: Templeton Foundation Press, 2008), p. 177.

[43]Ibid., 165.

[44]Ian G. Barbour, *Religion in an Age of Science: The Gifford Lectures 1989-1991, Vol. 1* (San Francisco, CA: Harper Collins, 1990).

[45]Ibid., xiv.

[46]James B. Ashbrook and Carol Rausch Albright, "Religion and Science Conversation: A Case Illustration," *Zygon: Journal of Religion and Science* 34 (September 1999): 399-418.

[47]Robert Wuthnow, "No Contradictions Here," in *The Religion and Science Debate: Why Does It Continue?* ed. Harold W. Attridge (New Haven, CT: Yale University Press, 2009), 161.

[48]Alvin Plantinga, "No Contradictions Here," in *The Religion and Science Debate: Why Does It Continue?* ed. Harold W. Attridge (New Haven, CT: Yale University Press, 2009), 94.

[49]Barbour, *Religion in an Age of Science: The Gifford Lectures 1989-1991, Vol. 1*, 21.